# Investigation of rare semileptonic $B_c \to (D_{s,d}^{(*)})\mu^+\mu^-$ decays with non-universal $Z'$ effect

**P. Maji**[*], **S. Mahata, P. Nayek, S. Biswas and S. Sahoo**[**]

Department of Physics, National Institute of Technology Durgapur

Durgapur-713209, West Bengal, India

*E-mail: majipriya@gmail.com, **E-mail: sukadevsahoo@yahoo.com

**Abstract**

In this paper, we analyze different decay observables of semileptonic decays $B_c \to (D_{s,d}^{(*)})\mu^+\mu^-$ such as branching ratio, forward-backward asymmetry, polarization fraction and lepton polarization asymmetry in the non-universal $Z'$ model. We have also studied the dependence of branching fraction to the new model parameters. Here, we have found that the values of different decay parameters increase in the $Z'$ model, which provides a possible way out for the search of new physics as well as the unknown phenomena of charm $B$ meson.

## I. Introduction

It is well-known that in spite of having enormous successful explanation of many experimental observations, the standard model (SM) carries some certain lacunae. Besides that, with little experimental validation, there remains a huge possibility of physics beyond the SM. In last few years some discrepancies have been observed in various meson decays, most notably in the angular observable $P_5'$ [1] of $B \to K^*\mu^+\mu^-$, branching ratio of $B \to \phi\mu^+\mu^-$ [2], lepton flavor non-universality parameter $R_{K^{(*)}}$ [3,4] and $R_{D^{(*)}}$ [5]. Due to the deficit in SM theory, these anomalies drive us to search for new physics (NP). High energy experiments at the LHC for the indirect search of rare decays of beauty and charm hadrons are dedicated to make precision measurements of the SM and beyond SM.

After the CLEO observation of $b \to s\gamma$ transition [6] rare decays of $B_{u,d,s}$ mesons become the main topic of interest. These studies would be more reliable if the results of $B_c$ meson discovered by the CDF Collaboration [7] are included. Experimentally, the CDF Collaboration found the $B_c$ meson in 1998 via semileptonic channel $B_c^+ \to J/\psi l^+ \nu$. The study of $B_c$ meson is itself quite exotic due to some outstanding features [8-10]. The $B_c$ meson [11] is composed of two heavy quarks $b$ and $c$, which are of different charge and flavor. Those heavy quarks are bound to the lowest state to form $B_c$ meson and thus several properties of its decay modes are different from other flavor neutral processes. The main difference between the weak decays of $B_c$ and $B_{u,d,s}$ is that the latter ones could be described in the background of heavy quark limit which provides some relations between the form factors of the physical process. But in the case of $B_c$ meson, heavy flavor and spin symmetries must be reconsidered as both the constituent quarks are heavy. Another important distinction between weak decays of $B_c$ meson associated with $b$ and $c$ quark decays includes significant difference of allowed kinematical region. The accessible kinematic range is broader in the decays of $B_c$ meson to charmonium and $D$ mesons than for the decays of $B_s$ and $B_d$ meson. As a result many weak decays are kinematically allowed in the former case but restricted in the latter one. Since other excited states of $\bar{b}c$ lie below the threshold of decay into the pair of $B$ and $D$ mesons, the strong and electromagnetic decay channels for these states are forbidden while the weak decays are allowed. $B_c$ meson persists more decay channels with a larger final phase space as the heavy quarks $b$ and $c$ can decay independently or both of them take part in a single process. The phase space for $c \to s$ transition is found to be smaller than that in $b \to c$ transition, but the CKM matrix element $|V_{cs}| \sim 1$ is much larger than $|V_{cb}| \sim 0.04$. Thus decay modes of $c$ quark provide dominant contribution (~70%) to the decay width of $B_c$ meson [12].

This meson offers a very rich laboratory for studying various decay channels which are essential for both theoretical and experimental aspects. With the possibility of an upcoming production of a large number of $B_c$ meson (about $10^8 \sim 10^{10}$) per year [13, 14] at future LHC run (with the luminosity values of $\mathcal{L} = 10^{34}$cm$^{-2}$s$^{-1}$ and $\sqrt{s} = 14$TeV), one might explore the rare semileptonic $B_c$ decays to $(D_s^{(*)}, D_d^{(*)})l^+l^-$ induced by single-quark flavor-changing neutral current (FCNC) $b \to s, d$ transitions. According to the GIM mechanism [15], these transitions are forbidden at tree level but allowed through electroweak loop diagrams. It indicates towards highly suppressed SM contribution. Due to this large suppression it is important to study these channels beyond the SM with new physics (NP) effects. This might provide useful probes to test the SM and detect the NP indirectly.

The processes induced by $b \to sl\bar{l}$ transition have been widely discussed in $B \to K^{(*)}l\bar{l}$ decays where $b \to sl\bar{l}$ has a dominant contribution. The annihilation diagrams are CKM suppressed $|V_{ub}^*V_{us}|/|V_{ts}^*V_{tb}| \sim \lambda^2$ and the spectator scattering is at the next order of $\alpha_s$. The CP violation in this channels is strongly suppressed in the SM due to the presence of only one independent CKM factor $V_{ts}^*V_{tb}$. On the other hand, for $b \to d$ modes all three CKM factors $V_{td}^*V_{tb}, V_{cd}^*V_{cb}, V_{ud}^*V_{ub}$ are of same order and hence it can induce notable CP-violating difference between the decay rates of $b \to dl^+l^-$ and $\bar{b} \to \bar{d}l^+l^-$. The theoretical investigation of these rare exclusive transitions could be done through two steps. Firstly, the effective Hamiltonian of these processes is calculated from leading and next-to-leading loop diagrams in the SM by using operator product expansion and renormalization group techniques. The reviews of this part are described in ref. [16, 17]. Secondly, the matrix elements of the effective Hamiltonian between hadronic states are needed. This part is model dependent as it requires nonperturbative QCD.

There are several approaches in literature where semileptonic $B_c$ decays have been investigated extensively. In ref. [18], authors described a detailed study of the exclusive semileptonic $B_c$ decays in the framework of Bauer-Stech-Wirbel. In refs. [19-21], the studies were done in the relativistic and/or constituent quark model, whereas in refs. [22, 23], $B_c \to D_s^*l^+l^-$ channels have been investigated in the SM with the fourth generation and supersymmetric models. In refs. [24, 25], the authors have presented the three-point QCD approach for their analysis. The light-front quark model was adopted by the authors in refs. [20, 26] for their needful probes. In ref. [27] authors have explored the perturbative QCD approach to study the semileptonic $B_c$ decay channels. New physics contributions to $B_c \to D_s^*l^+l^-$ decay has been studied extensively in single universal extra dimension model [28] and also analyzed in a model-independent way using an effective Hamiltonian approach [29, 30].

In this work, we have taken the QCD-motivated relativistic quark background and supplement the previous analysis of different decay observables of $B_c \to (D_s^{(*)}, D_d^{(*)})l^+l^-$ by considering the effect of non-universal $Z'$ boson. In the relativistic quark model, quasipotential approach has been considered where a meson is described as a bound state with a wave function consisting of the solution of Schrodinger type quasipotential equation. This model provides particular attention to the inclusion of negative-energy contributions and the relativistic transformation of the wave function from the rest to moving reference frame. The numerical calculations are based on these relativistic wave functions which are obtained previously from meson mass spectra. Another advantage of this approach is that the electroweak matrix elements between meson states with a consistent relativistic effect allow to determine the form factor dependence on the momentum transfer. This dependence is reliable in whole accessible kinematic range without using any *ad hoc* assumption and extrapolation. The form factors have been expressed as overlap integrals of the meson wave function. Here one has to check the fulfillment of model-independent symmetry relations among the form factors arising in heavy quark and large energy limits are fulfilled. We have followed the calculation of ref. [31] for the values of $B_c \to D_s^{(*)}, D_d^{(*)}$ form factors. This paper is organized as follows. In section II, the formalism of effective Hamiltonian for $B_c \to (D_s^{(*)}, D_d^{(*)})l^+l^-$ decay modes have been presented. In section III, different decay observables for the above processes are given in terms of helicity amplitudes. In section IV, the outline of the non-universal $Z'$ model has been given. In section V, we have analyzed our predicted results. Section VI consists of a summary and concluding remarks.

## II. Formalism of Effective Hamiltonian

Usually rare B decays are described by low energy effective Hamiltonian obtained by integrating out the heavy degrees of freedom of the top quark and $W$ boson. Short-distance contributions contained in Wilson coefficients are separated by operator product expansion and calculated perturbatively. Long-distance contributions are contained in matrix elements of local operators which are calculated in a non-perturbative approach.

The effective Hamiltonian for $b \to ql^+l^-$ (where $q = s, d$) transition renormalized at a scale $\mu \approx m_b$ is given by [32]

$$\mathcal{H}_{eff} = -\frac{4G_F}{\sqrt{2}} V_{tq}^* V_{tb} \sum_{i=1}^{10} C_i \mathcal{O}_i, \quad (1)$$

where $G_F$ is the Fermi constant, $V_{tj}$ are CKM matrix elements, $C_i$ are the Wilson coefficients and $\mathcal{O}_i$ are the standard model operator basis which could be found in [16]. $\mathcal{O}_i(\mu)$ ($i = 1, … … … 6$) represent the four-quark operators, $i = 7, 8$ are for dipole operators and $i = 9, 10$ represent semileptonic electroweak operators. Here, the operators $\mathcal{O}_7, \mathcal{O}_9$ and $\mathcal{O}_{10}$ are mainly responsible for these decay modes. From the reduced effective Hamiltonian we can get the free quark decay amplitude which is written as

$$\mathcal{M}(b \to ql^+l^-) = \frac{G_F \alpha_{em}}{2\sqrt{2}\pi} V_{tb} V_{ts}^* \{C_9^{eff}(\mu) \bar{q} \gamma_\mu (1-\gamma_5) b (\bar{l}\gamma^\mu l) + C_{10}(\mu) \bar{q} \gamma_\mu (1-\gamma_5) b (\bar{l}\gamma^\mu \gamma_5 l)$$

$$-\frac{2m_b}{q^2} C_7^{eff}(\mu)\bar{q}i\sigma_{\mu\nu}q^\nu(1+\gamma_5)b(\bar{l}\gamma^\mu l)\} \qquad (2)$$

where, $\alpha_{em}$ is the fine structure constant. Within SM $C_7^{eff}$ in leading logarithm approximation is written as [33]

$$C_7^{eff}(\mu) = \eta^{\frac{16}{23}} C_7(m_W) + \frac{8}{3}\left(\eta^{\frac{14}{23}} - \eta^{\frac{16}{23}}\right) C_8(m_W) + C_2(m_W) \sum_{i=1}^{8} h_i \eta^{a_i}, \qquad (3)$$

where, $C_2(m_W) = 1$ and $C_7(m_W), C_8(m_W)$ are given in ref. [34]. The coefficients $a_i$ and $h_i$ are given as [35, 17],

$a_i =$ (14/23, 16/23, 6/23, -12/23, 0.4086, -0.4230, -0.8994, 0.1456),
$h_i =$ (2.2996, -1.0880, -3/7, -1/14, -0.6494, -0.0380, -0.0186, -0.0057).

The parameter $\eta$ in eq. (7) is defined as, $\eta = \frac{\alpha_s(\mu_W)}{\alpha_s(\mu_b)}$.

$C_9^{eff}$ contains short-distance perturbative contribution and long-distance contribution terms. Within SM $C_9^{eff}$ is written as

$$C_9^{eff} = C_9 + y_{pert}(q^2) + y_{BW}(q^2), \qquad (4)$$

where, $q^2$ is the four-momentum squared of the lepton pair. The short-distance contribution (perturbative part) denoted by $y_{pert}(q^2)$ [31] involves the indirect contributions coming from the matrix element of four quark operators. The long-distance part denoted by $y_{BW}(q^2)$ have $c\bar{c}$ intermediate states, i. e. $J/\psi$ family [36]. By introducing the Breit-Wigner formula the explicit expression of $y_{BW}(q^2)$ is parameterized [30] and is given in appendix A. $c\bar{c}$ resonances provide a large peak in the decay distribution due to which hadronic uncertainties are coming to the semileptonic decay modes. To apply these relations in $B_c$ decay modes we need to find the matrix elements of the operators $\bar{q}\gamma_\mu(1-\gamma_5)b$ and $\bar{q}\sigma_{\mu\nu}q^\nu(1+\gamma_5)b$ between initial and final hadronic states which is based on a non-perturbative approach.

The long-distance processes considered here are induced by resonance cascade modes such as $B_c \to D_{s(d)}^{(*)} V \to D_{s(d)}^{(*)} l\bar{l}$. The contributions of these transitions could be termed after the relationship $Br(B_c \to D_{s(d)}^{(*)} l\bar{l})_{cascade} \sim Br(B_c \to D_{s(d)}^{(*)} V) \times Br(V \to l\bar{l})$. The resonances $V$ denote $J^{PC} = 1^{--}$ mesons which could be $\bar{u}u, \bar{d}d, \bar{s}s$ and $\bar{c}c$ bound states. In our analysis we neglect the effects of the $B_c \to D_{s(d)}^{(*)} \rho(\omega, \phi)$ cascade decays. Due to the Okubo-Zweig-Iizuka (OZI) rules allow the strong decays of $\rho, \omega$ and $\phi$ mesons, while the decay modes of $J/\psi(\psi(2S))$ are suppressed by OZI rules. So the transitions $\rho(\omega, \phi) \to l\bar{l}$ induced by electromagnetic interaction are of smaller branching fraction than the processes $J/\psi(\psi(2S)) \to l\bar{l}$. On the other side, $B_c \to D_s^{(*)} \rho(\omega, \phi)$ modes are suppressed because of small CKM matrix elements $V_{ub}$ and $V_{us}$. Due to this reason Wilson coefficients $C_{3-6}$ are also small providing lesser branching fraction of $B_c \to D_{s,d}^{(*)} \rho(\omega, \phi)$. So here we have considered only $B_c \to D_{s(d)}^{(*)} J/\psi(\psi(2S)) \to D_{s(d)}^{(*)} l\bar{l}$ processes [37, 38].

# III. Decay observables of $B_c \to D_{s,d}^{(*)} l\bar{l}$ processes

In this section we have presented the explicit expressions of different decay observables of the semileptonic decay channels $B_c \to (D_{s,d}^{(*)})l^+l^-$. The matrix elements could be parameterized in terms of different hadronic form factors and are given in appendix A. The obtained form factors are consistent with all model independent symmetry relation [39, 40] within the limit of infinitely heavy quark mass and large energy of the final meson. For the helicity amplitudes we recall the techniques of ref. [41, 42] followed the ref. [30]. These amplitudes are given in appendix A. The subscripts $\pm, 0, t$ denote transverse, longitudinal and time helicity components. As the final meson $D_{s,d}$ are pseudo-scalar mesons and do not have any polarization direction, so the transverse helicity amplitudes for $B_c \to D_{s,d} l^+l^-$ channels are 0.

Based on the calculation of ref. [30, 37], the three-body $B_c \to D_{s(d)} l^+l^-$ and $B_c \to D_{s(d)}^* l^+l^-$ differential decay rate are given by,

$$\frac{d\Gamma_{s(d)}}{dq^2} = \frac{G_F^2}{(2\pi)^3}\left(\frac{\alpha_e |V_{tb}V_{ts(td)}^*|}{2\pi}\right)^2 \frac{\lambda^{1/2}q^2}{48 M_{B_c}^3}\sqrt{1-\frac{4m_l^2}{q^2}}\left[H^{(1)}H^{\dagger(1)}\left(1+\frac{4m_l^2}{q^2}\right)\right.$$
$$\left. + H^{(2)}H^{\dagger(2)}\left(1-\frac{4m_l^2}{q^2}\right) + \frac{2m_l^2}{q^2}3H_t^{(2)}H_t^{\dagger(2)}\right], \quad (5)$$

where $m_l$ is the lepton mass and

$$H^{(i)}H^{\dagger(i)} = H_+^{(i)}H_+^{\dagger(i)} + H_-^{(i)}H_-^{\dagger(i)} + H_0^{(i)}H_0^{\dagger(i)}. \quad (6)$$

Besides that we also study some other observables like forward-backward asymmetry ($A_{FB}$) and the longitudinal polarization fraction ($P_L$) of the final vector meson in the decay $B_c \to D_{s,d}^* l^+l^-$. While analyzing the channel $B \to K^* l^+l^-$, $A_{FB}$ and $P_L$ have got wide attention both theoretically and experimentally. It is expected to collect more information on the Wilson coefficient by investigating these observables. The forward-backward asymmetry ($A_{FB}$) is given by [30]

$$A_{FB}(q^2) = \frac{3}{4}\sqrt{1-\frac{4m_l^2}{q^2}}$$
$$\left\{\frac{Re\left(H_+^{(1)}H_+^{\dagger(2)}\right) - Re(H_-^{(1)}H_-^{\dagger(2)})}{H^{(1)}H^{\dagger(1)}\left(1+\frac{4m_l^2}{q^2}\right) + H^{(2)}H^{\dagger(2)}\left(1-\frac{4m_l^2}{q^2}\right) + \frac{2m_l^2}{q^2}3H_t^{(2)}H_t^{\dagger(2)}}\right\}. \quad (7)$$

A notable fact is that forward-backward asymmetry observable for the $B_c \to D_{s,d} l^+l^-$ channel is zero in the SM which consequently states parity-even nature. The non-zero value of $A_{FB}$ indicates parity-odd effects arising due to parity-conserving contribution coming from scalar-vector interference. $A_{FB} \neq 0$ might be possible if it receives contribution from scalar, pseudoscalar or tensor new physics operator. But in our model no new operator has been introduced instead only the Wilson coefficients have been modified. So we stick to the zero forward backward asymmetry and do not discuss this observable for $B_c \to D_{s,d} l^+l^-$.

Similarly, the longitudinal polarization fraction ($P_L$) of the $D^*_{s(d)}$ meson is written as [30]

$$P_L(q^2) = \frac{H_0^{(1)}H_0^{\dagger(1)}\left(1+\frac{4m_l^2}{q^2}\right) + H_0^{(2)}H_0^{\dagger(2)}\left(1-\frac{4m_l^2}{q^2}\right) + \frac{2m_l^2}{q^2}3H_t^{(2)}H_t^{\dagger(2)}}{H^{(1)}H^{\dagger(1)}\left(1+\frac{4m_l^2}{q^2}\right) + H^{(2)}H^{\dagger(2)}\left(1-\frac{4m_l^2}{q^2}\right) + \frac{2m_l^2}{q^2}3H_t^{(2)}H_t^{\dagger(2)}}. \qquad (8)$$

Here, we only investigate the longitudinal polarization of the final vector meson. The transverse polarizations $P_T$ could be obtained from the relation $P_T = 1 - P_L$. Furthermore, the leptonic polarization asymmetry ($A_{P_L}$) is defined as [37],

$$A_{P_L} = \frac{\frac{dBr_{h=-1}}{dq^2} - \frac{dBr_{h=1}}{dq^2}}{\frac{dBr_{h=-1}}{dq^2} + \frac{dBr_{h=1}}{dq^2}}$$

$$= \sqrt{1-\frac{4m_l^2}{q^2}}\frac{2[Re\left(H_+^{(1)}H_+^{\dagger(2)}\right) + Re\left(H_-^{(1)}H_-^{\dagger(2)}\right) + Re\left(H_0^{(1)}H_0^{\dagger(2)}\right)]}{H^{(1)}H^{\dagger(1)}\left(1+\frac{4m_l^2}{q^2}\right) + H^{(2)}H^{\dagger(2)}\left(1-\frac{4m_l^2}{q^2}\right) + \frac{2m_l^2}{q^2}3H_t^{(2)}H_t^{\dagger(2)}} \qquad (9)$$

## IV. The non-universal $Z'$ model

There are several models beyond the SM which predict the existence of exotic fermions. If the new exotic fermions have different $U(1)'$ charges as in $E_6$ models [42-44], mixing between ordinary (doublet) and exotic singlet left-handed fermions induces undesirable FCNC mediated by the SM $Z$ boson. In contrast, the mixing between right handed ordinary and exotic fermions induces FCNC mediated by $Z'$ boson.

Here, the choice of the non-universal $Z'$ model [45-48] is considered to be the most economical as it requires one extra $U(1)'$ gauge symmetry associated with a neutral gauge boson called $Z'$ boson. Basic formalism of the family non-universal $Z'$ model with FCNCs can be found in [46, 49, 50]. The main attraction of this model is that the FCNC transitions could occur at tree level due to the off-diagonal (flavor changing) couplings of non-universal $Z'$ with fermions, which is not allowed under SM consideration. Various studies of the non-universal $Z'$ model have been done assuming diagonal as well as vanishing right-handed quark couplings with $Z'$ boson. It is observed that it can help to resolve the puzzles of rare B meson decays such as $B - \bar{B}$ mixing phase [51], $\pi - K$ puzzle [50, 52], $\pi - \pi$ puzzle [53, 54], etc.

In this model the $Z'$ part of the neutral-current Lagrangian within the basis of gauge eigenstates of all fields is written as

$$\mathcal{L}^{Z'} = -g'J'_\mu Z'^\mu \qquad (10)$$

where $g'$ is the new gauge coupling of the $U(1)'$ group at the $M_W$ scale.

The $U(1)'$ currents for $Z'$ boson in the appropriate gauge basis is

$$J'_\mu = \sum_{i,j} \bar{\psi}_i \gamma_\mu \left[\epsilon_{\psi_{L_{ij}}} P_L + \epsilon_{\psi_{R_{ij}}} P_R\right]\psi_j, \qquad (11)$$

where $i$ is the family index and $\psi$ stands for the fermions (up- or down-type quarks, or charged or neutral leptons). $P_{L,R} = (1 \mp \gamma_5)/2$ and $\epsilon_{\psi_{R,L_{ij}}}$ represent the chiral couplings of $Z'$ boson. The chiral $Z'$ coupling matrices in the fermion mass eigenstate basis are given as

$$B_{ij}^{\psi_L} \equiv \left(V_L^{\psi} \epsilon_{\psi_L} V_L^{\psi\dagger}\right)_{ij}, B_{ij}^{\psi_R} \equiv \left(V_R^{\psi} \epsilon_{\psi_R} V_R^{\psi\dagger}\right)_{ij}. \tag{12}$$

These couplings may contain CP-violating phases beyond the SM. As long as the $\epsilon$ matrices are not proportional to the identity, the $B$ matrices will have non-zero off-diagonal elements that induce FCNC interaction at tree level. We have chosen the basis such as $\epsilon_{\psi_R} \propto I$, so that the right-handed couplings are vanished within this framework. If $B_{ij}^{\psi_R}$ is non-diagonal different chirality structures will be induced in $B$ decays which generate new operators to the effective Hamiltonian. The presence of new chirally flipped operators might treat these transitions differently and may produce deviations from the SM. But those discussions are beyond this paper as we only modify the Wilson coefficients in our analysis and not create any new operators except the SM semileptonic operators.

For $\to q$ ($q = s, d$) transition, the $Z'bq$ couplings are generated as [55],

$$\mathcal{L}_{FCNC}^{Z'} = -\text{g}'\left(B_{sb}^L \bar{s}_L \gamma_\mu b_L + B_{sb}^R \bar{s}_R \gamma_\mu b_R\right) Z'^\mu + h.c. \tag{13}$$

The effective Hamiltonian for the above transition mediated by $Z'$ boson can be written as

$$H_{eff}^{Z'} = \frac{8G_F}{\sqrt{2}} \left(\rho_{sb}^L \bar{s}_L \gamma_\mu b_L + \rho_{sb}^R \bar{s}_R \gamma_\mu b_R\right)\left(\rho_{ll}^L \bar{l}_L \gamma_\mu l_L + \rho_{ll}^R \bar{l}_R \gamma_\mu l_R\right) \tag{14}$$

where,

$$\rho_{ff'}^{L,R} \equiv \frac{\text{g}' M_Z}{\text{g} M_{Z'}} B_{ff'}^{L,R}. \tag{15}$$

The value of $\left|\frac{\text{g}'}{\text{g}}\right|$ is not determined yet. But, it is expected that $\left|\frac{\text{g}'}{\text{g}}\right| \sim 1$ as both $U(1)$ groups are coming from the same GUT. Throughout the whole analysis, we ignore the renormalization group running effects due to these new contributions. To avoid too many free parameters, we assume that the FCNC couplings of the $Z'$ and quarks only occur in the left-handed sector. Therefore, $\rho_{sb}^R = 0$ and the effects of the $Z'$ FCNC currents simply modify the Wilson coefficients $C_9$ and $C_{10}$. Since the $Z'$ boson has not yet been discovered, its mass is unknown. But, there are stringent limits on the mass of an extra $Z'$ boson obtained by CDF, DØ and LEP 2, and on the $Z - Z'$ mixing angle $\theta_{ZZ'}$ [56]. The precision electroweak (EW) data strongly constrain on $\theta_{ZZ'}$ to be very small such as $|\theta_{ZZ'}| \leq 8.1 \times 10^{-3}$. Using the current LHC Drell-Yan data, authors of ref. [57-59] obtained the lower limit of mass of $Z'$ as $M_{Z'} > 4.4$ TeV. Recently, in ref. [60] the constraints on the mixing angle $\theta_{ZZ'}$ has been derived from resonant diboson searches at the LHC at $\sqrt{s} = 13$ TeV which is of the order of a few $\times 10^{-4}$. Due to the tininess of $\theta_{ZZ'}$ we can neglect the $Z - Z'$ mixing and consider that the couplings of only the right-handed quarks with $Z'$ are diagonal. Now we can write the effective Hamiltonian for the transition $b \to q l^+ l^-$ mediated by $Z'$ FCNC as

$$H_{eff}^{Z'} = \frac{2G_F}{\sqrt{2}\pi} V_{tb} V_{tq}^* \left[ \frac{B_{qb}^L S_{ll}^L}{V_{tb} V_{tq}^*} \bar{q}\gamma_\mu(1-\gamma_5)b\bar{l}\gamma^\mu(1-\gamma_5)l \right.$$
$$\left. + \frac{B_{qb}^L S_{ll}^R}{V_{tb} V_{tq}^*} \bar{q}\gamma_\mu(1-\gamma_5)b\bar{l}\gamma^\mu(1+\gamma_5)l \right], \quad (16)$$

where $B_{qb}^L = |B_{qb}^L|e^{-i\varphi_{qb}}$ represents the off-diagonal left-handed couplings of $Z'$ boson with the quark sector and $\varphi_{qb}$ is the new weak phase angle. Now the concise effective Hamiltonian is given by [61]

$$H_{eff}^{Z'} = -\frac{4G_F}{\sqrt{2}} V_{tb} V_{ts}^* [\Lambda_{sb} C_9^{Z'} O_9 + \Lambda_{sb} C_{10}^{Z'} O_{10}], \quad (17)$$

where
$$\Lambda_{sb} = \frac{4\pi e^{-i\varphi_{sb}}}{\alpha V_{tb} V_{td}^*},$$
$$C_9^{Z'} = |B_{sb}| S_{LL},$$

and
$$C_{10}^{Z'} = |B_{sb}| D_{LL}. \quad (18)$$

Here, $S_{LL} = S_{ll}^L + S_{ll}^R$ and $D_{LL} = S_{ll}^L - S_{ll}^R$.

The terms $S_{ll}^L$ and $S_{ll}^R$ denote the couplings of $Z'$ boson with left- and right-handed leptons respectively. The numerical values of the $Z'$ couplings suffer from several constraints that arise due to different exclusive and inclusive B decays [51, 62]. We have used two scenarios as described in Table 1 in our calculation, corresponding to different fitting values of $B_s - \bar{B}_s$ and $B_d - \bar{B}_d$ mixing data that present the couplings as well as the weak phase angle. The values of input parameters of $|B_{sb}|$ and $\phi_{sb}$ are set by UTfit collaborations [63], whereas $|B_{db}|$ and $\phi_{db}$ are recollected from ref. [64].

Table 1: Input parameters for non-universal $Z'$ model [65, 66]

|  | $|B_{sb}| \times 10^{-3}$ | $\phi_{sb}$(Degree) | $|B_{db}| \times 10^{-3}$ | $\phi_{db}$(Degree) | $S_{LL} \times 10^{-2}$ | $D_{LL} \times 10^{-2}$ |
|---|---|---|---|---|---|---|
| $S_1$ | 1.09 ± 0.22 | −72 ± 7 | 0.16±0.08 | −33±45 | −2.8 ± 3.9 | −6.7 ± 2.6 |
| $S_2$ | 2.20 ± 0.15 | −82 ± 4 | 0.19±0.05 | −50±20 | −1.2 ± 1.4 | −2.5 ± 0.9 |

The contribution of non-universal $Z'$ boson to the branching ratio, FB asymmetry and lepton polarization asymmetry described in the next section are considered only for muonic channels i.e. $B_c \to D_{s,d}^{(*)} \mu^+ \mu^-$ decay modes.

### V. Numerical analysis

In this paper, we have analyzed different decay observables like branching ratio, forward-backward asymmetry, polarization fraction and lepton polarization asymmetry in the non-universal $Z'$ model. The pictorial descriptions provide a good knowledge about the deviation of these observables in the NP model from their SM predictions. We have plotted all these parameters with the variation of total momentum transfer squared i.e. $q^2$ and their graphical representations could be found below. We have also plotted the dependence of branching ratio on the NP model parameters $D_{LL}$ and $S_{LL}$ with different $\phi_{sb}$. Our whole study could lead us to the following consequences.

(i) From the illustrations of branching fraction (Fig. 9-12) of all those decay above some noticeable deviations from their SM values are found in low recoil, i.e., high $q^2$ region. Figs. 1-8 depict the dependence of branching fraction to the model parameters. Here we have taken the central values of $|B_{sb}|$ and $|B_{db}|$ for both the scenarios $S_1$ and $S_2$ and the value of new weak phase angle $\phi_{sb}$ and $\phi_{db}$ are changed accordingly. For $B_c \to (D_s, D_s^*)\mu^+\mu^-$ (Fig. 1 and Fig. 3) it is found that $\phi_{sb} = -65°$ maximally increases the value of branching fraction in the first scenario $S_1$. It is also seen from Fig. 1 (A) and Fig. 3 (A) that for smaller $D_{LL}$ the branching fraction gets increased whereas a saturated dependency on $S_{LL}$ could be seen from Fig. 1 (B) and Fig. 3 (B). In scenario $S_2$ the branching fraction depends on $D_{LL}$, $S_{LL}$ and $\phi_{sb}$ almost in similar fashion (Fig. 2 and Fig. 4). From Figs. 5 (A), 6 (A), 7 (A) and 8 (A), it is clear that the branching ratio of $B_c \to (D_d, D_d^*)\mu^+\mu^-$ directly depend on $\phi_{db}$ but inversely depend on $D_{LL}$. On the other hand, Fig. 5 (B) and 7 (B) also show direct dependence of branching ratio on $S_{LL}$ and Figs. 6 (B) and 7 (B) show the same nature but below the cuts: $S_{LL} = 0.03$ for $B_c \to D_d\mu^+\mu^-$ and $S_{LL} = 0.04$ for $B_c \to D_d^*\mu^+\mu^-$.

(ii) Fig. 13 and Fig. 14 show the variation of forward-backward asymmetry with $q^2$. This observable is a quite interesting property for any decay channel as it is sensitive to parity status of any interaction. At low $q^2$ region, parity conserving photonic interaction is relatively dominant leading to a small FB asymmetry. But in higher momentum region (i.e., large $q^2$), parity-violating Z- and W-boson contributions become more significant. As a consequence FB asymmetry becomes larger. For $B_c \to D_s^*\mu^+\mu^-$ the zero crossing is shifted to 3.2 GeV$^2$ from 2.1 GeV$^2$ in NP model and both the scenarios overlap with each other and lie below the SM. But there is no such shift of zero-crossing found for $B_c \to D_d^*\mu^+\mu^-$ channel and also $S_2$ goes little above from the SM whereas $S_1$ stays below it.

(iii) Fig. 15 and Fig. 16 depict the polarization fraction of $B_c \to (D_s^*, D_d^*)\mu^+\mu^-$ decay channels and it is found that no NP contribution is noticeable for this observable.

(iv) In Figs. 17-20 the lepton polarization asymmetry for all the four decay modes is presented. In the SM $A_{P_L}$ is $\sim -0.1$. In non-universal $Z'$ model the lepton polarization asymmetry of $B_c \to (D_s, D_s^*)\mu^+\mu^-$ have got a noticeable increment from -0.1 for both the scenarios $S_1$ and $S_2$. For $B_c \to (D_d, D_d^*)\mu^+\mu^-$ decay channels we have found a significant increment for $S_1$ but comparably lesser increase for $S_2$. An interesting fact came out in this study that the lepton longitudinal polarization asymmetry for $B_c \to D_{s,d}^*\mu^+\mu^-$ got a positive value in very low $q^2$ region. This phenomenon could be an ideal probe to investigate the spin direction of final state leptons in vector meson decay.

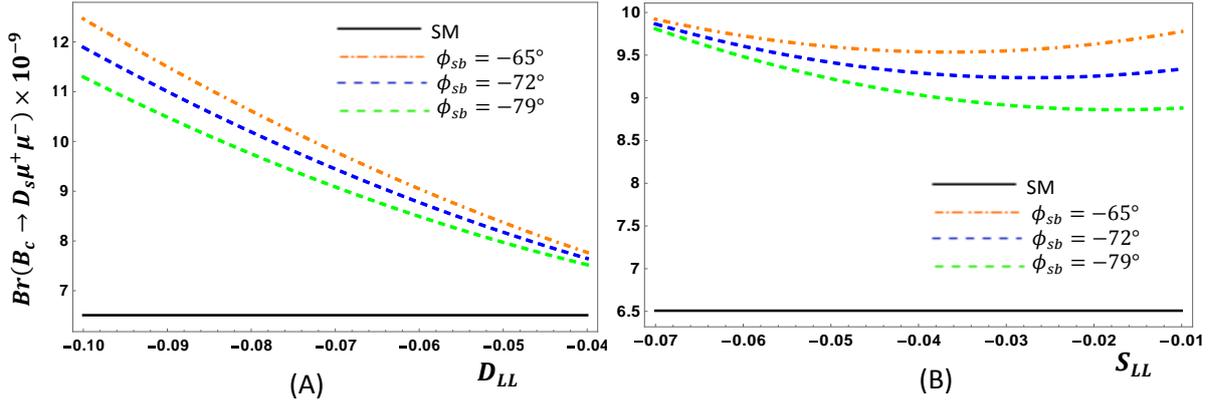

Fig. 1: The dependence of $Br(B_c \to D_s \mu^+ \mu^-)$ on $D_{LL}$ and $\phi_{sb}$ is shown in (A) whereas on $S_{LL}$ and $\phi_{sb}$ is shown in (B) for $\mathcal{S}_1$.

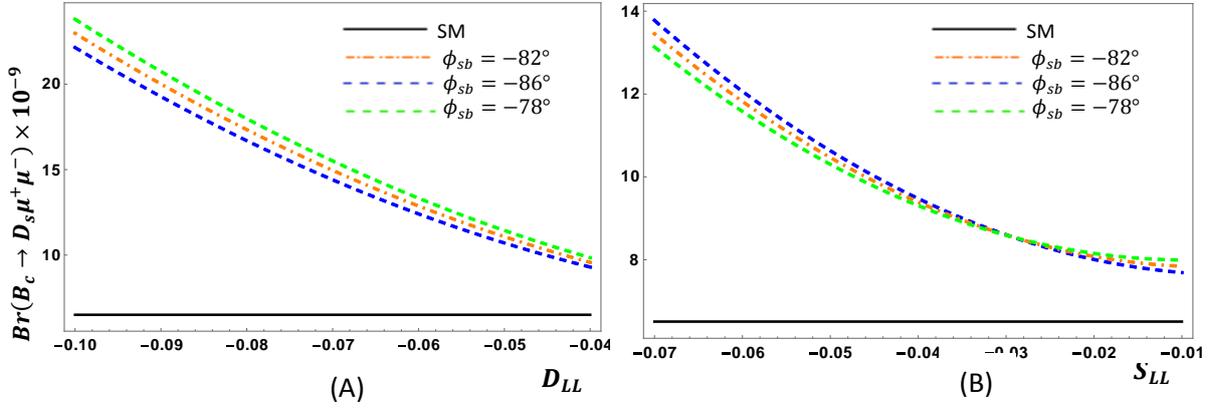

Fig. 2: The dependence of $Br(B_c \to D_s \mu^+ \mu^-)$ on $D_{LL}$ and $\phi_{sb}$ is shown in (A) whereas on $S_{LL}$ and $\phi_{sb}$ is shown in (B) for $\mathcal{S}_2$.

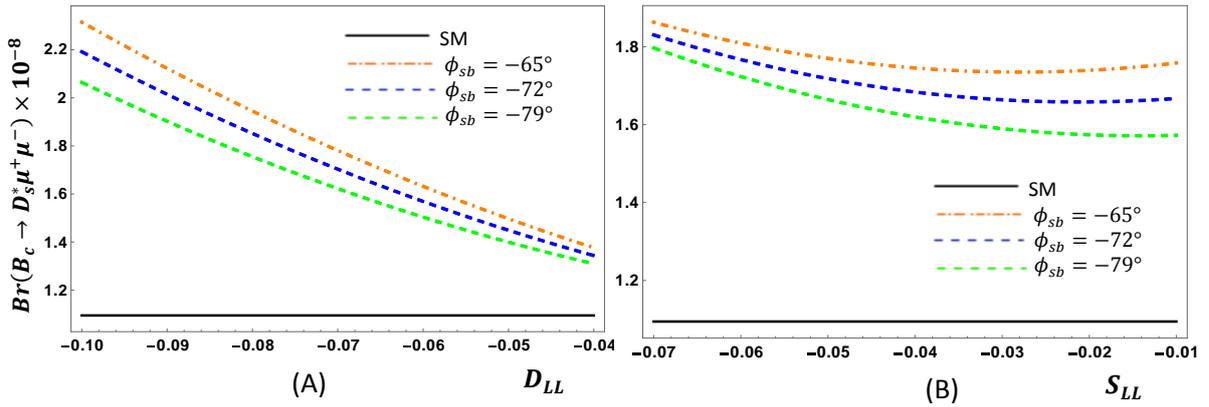

Fig. 3: The dependence of $Br(B_c \to D_s^* \mu^+ \mu^-)$ on $D_{LL}$ and $\phi_{sb}$ is shown in (A) whereas on $S_{LL}$ and $\phi_{sb}$ is shown in (B) for $\mathcal{S}_1$.

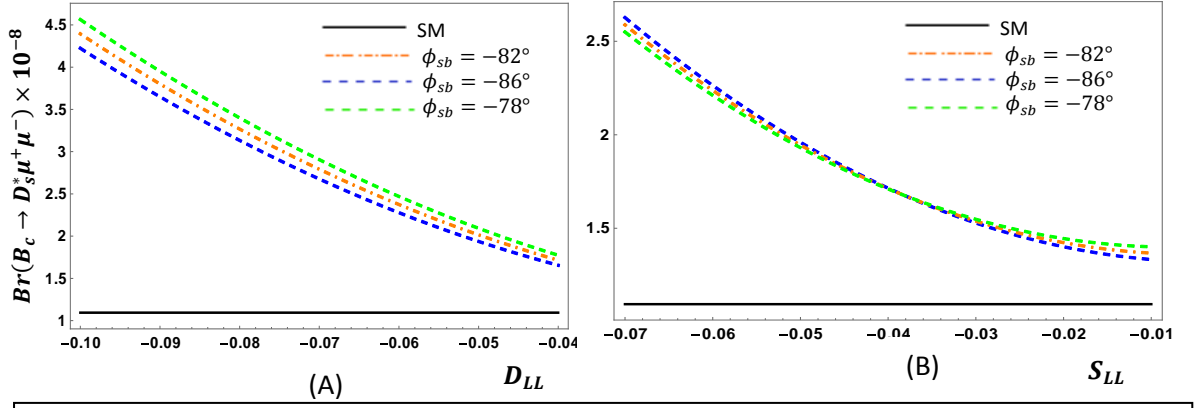

Fig. 4: The dependence of $Br(B_c \to D_s^* \mu^+ \mu^-)$ on $D_{LL}$ and $\phi_{sb}$ is shown in (A) whereas on $S_{LL}$ and $\phi_{sb}$ is shown in (B) for $\mathcal{S}_2$.

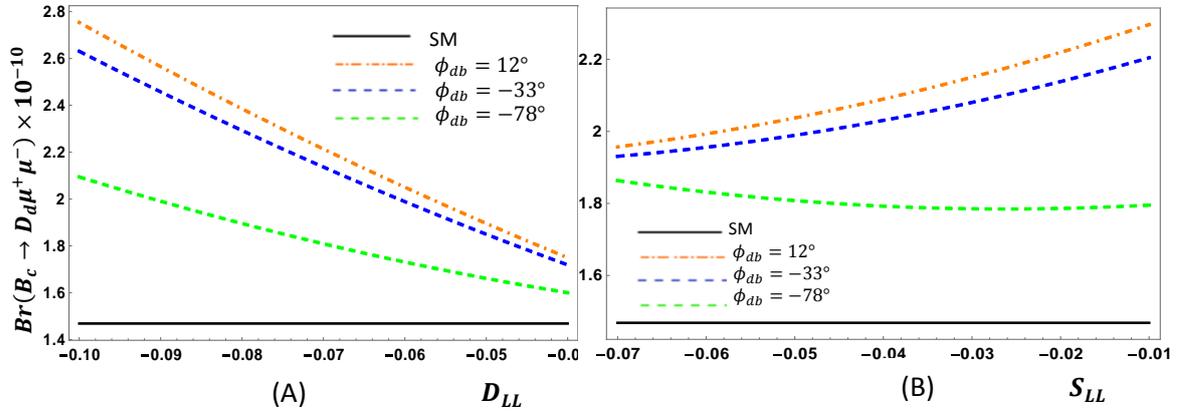

Fig. 5: The dependence of $Br(B_c \to D_d \mu^+ \mu^-)$ on $D_{LL}$ and $\phi_{db}$ is shown in (A) whereas on $S_{LL}$ and $\phi_{db}$ is shown in (B) for $\mathcal{S}_1$.

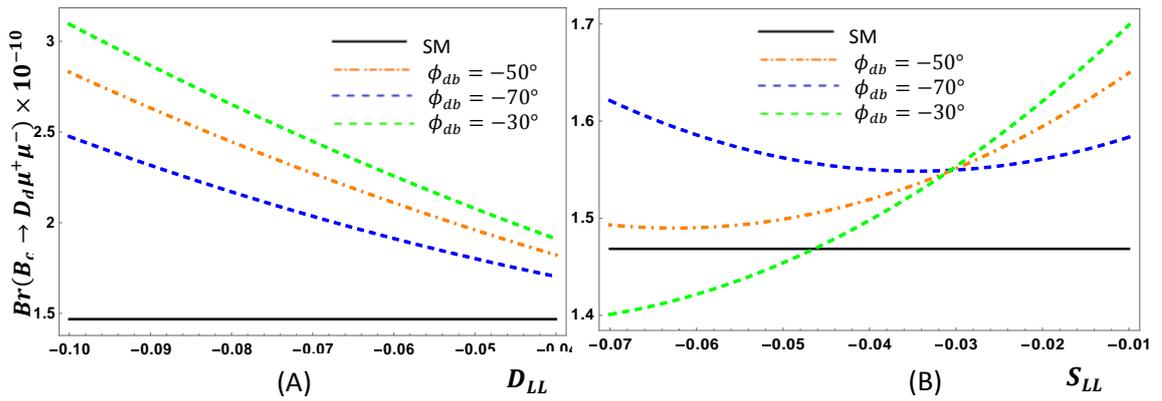

Fig. 6: The dependence of $Br(B_c \to D_d \mu^+ \mu^-)$ on $D_{LL}$ and $\phi_{db}$ is shown in (A) whereas on $S_{LL}$ and $\phi_{db}$ is shown in (B) for $\mathcal{S}_2$.

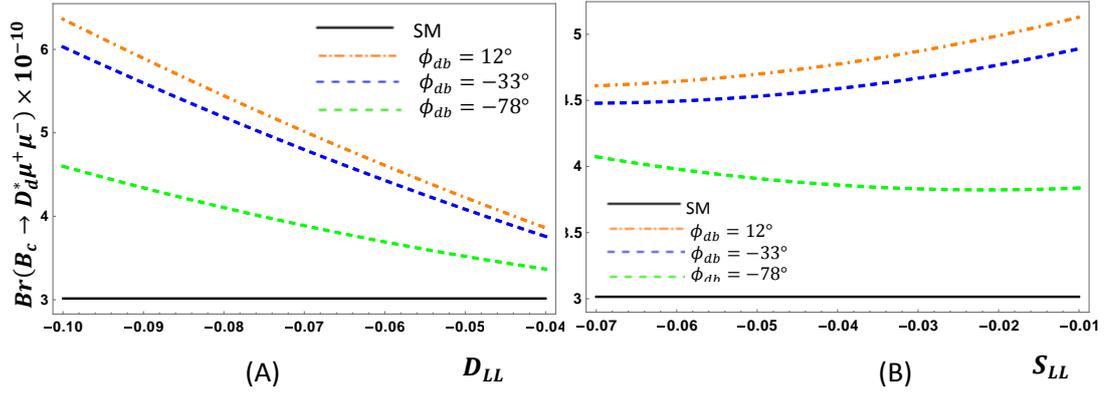

Fig. 7: The dependence of $Br(B_c \to D_d^* \mu^+ \mu^-)$ on $D_{LL}$ and $\phi_{db}$ is shown in (A) whereas on $S_{LL}$ and $\phi_{db}$ is shown in (B) for $\mathcal{S}_1$.

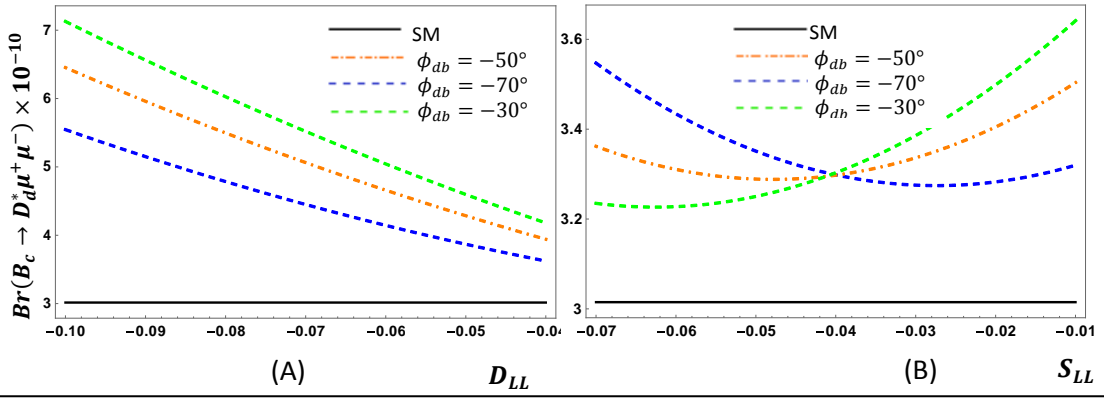

Fig. 8: The dependence of $Br(B_c \to D_d^* \mu^+ \mu^-)$ on $D_{LL}$ and $\phi_{db}$ is shown in (A) whereas on $S_{LL}$ and $\phi_{db}$ is shown in (B) for $\mathcal{S}_2$.

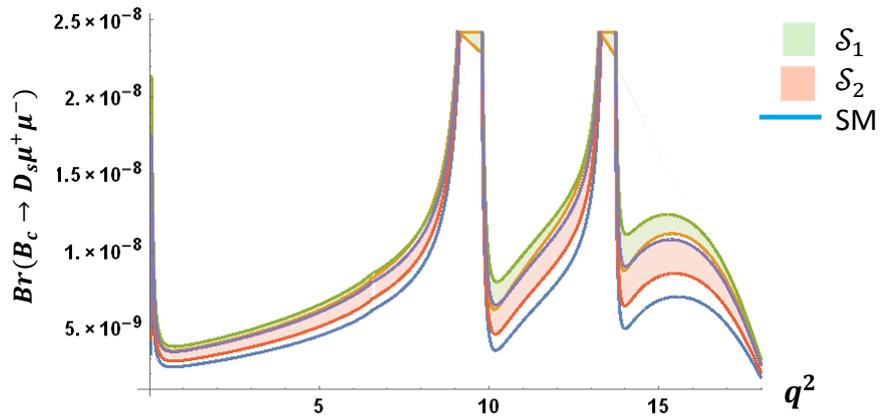

Fig. 9: The variation of branching fraction of $B_c \to D_s \mu^+ \mu^-$ with $q^2$.

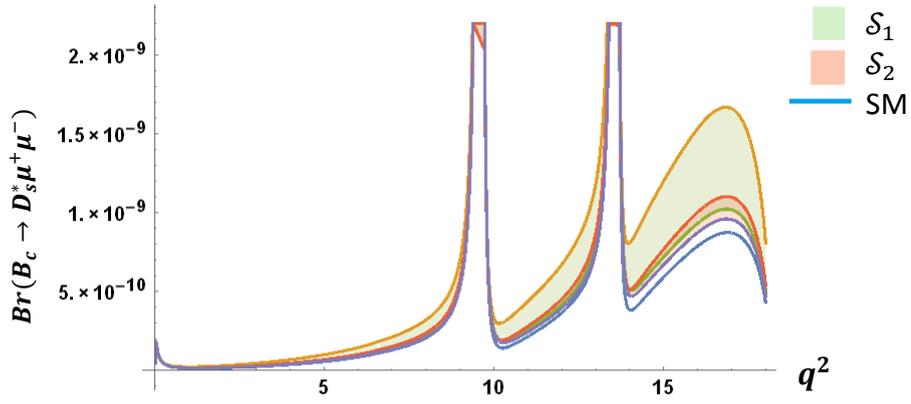

Fig. 10: The variation of branching fraction of $B_c \to D_s^* \mu^+ \mu^-$ with $q^2$.

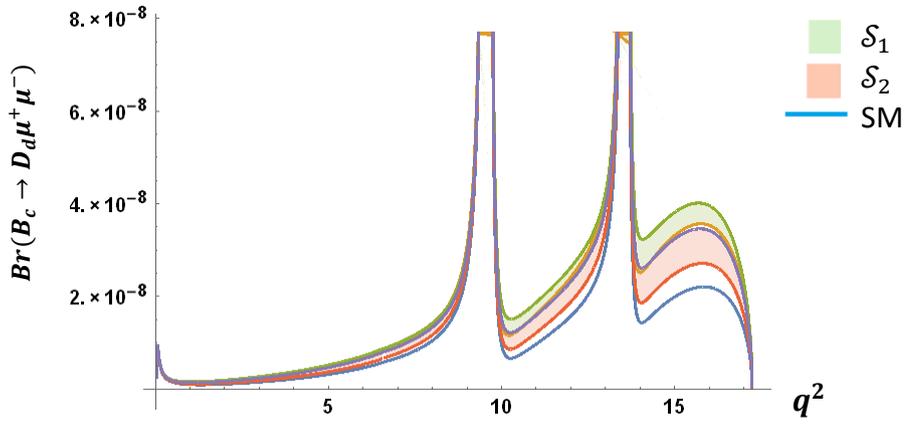

Fig. 11: The variation of branching fraction of $B_c \to D_d \mu^+ \mu^-$ with $q^2$.

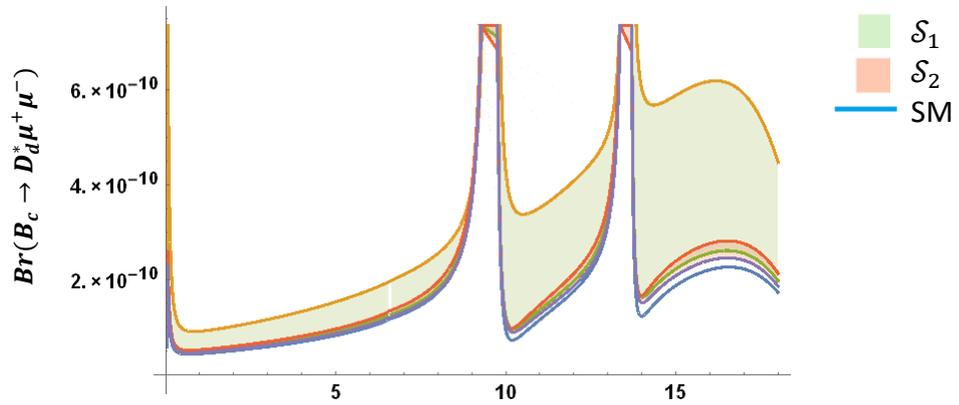

Fig. 12: The variation of branching fraction of $B_c \to D_d^* \mu^+ \mu^-$ with $q^2$.

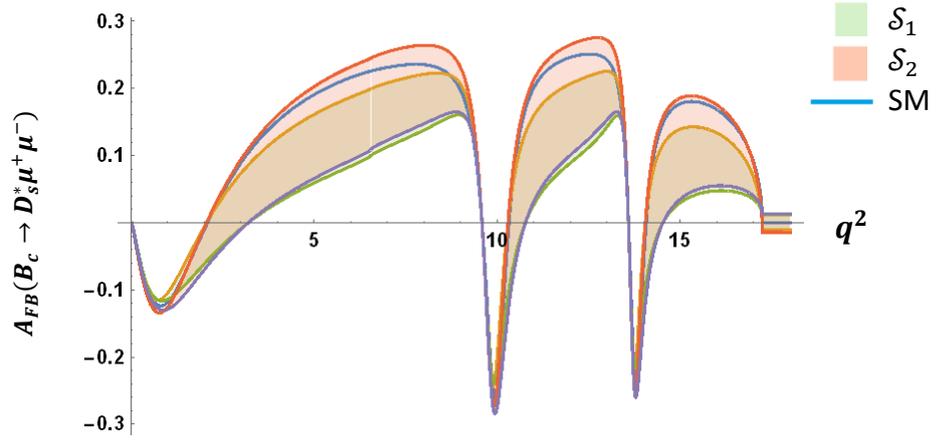

Fig. 13: The variation of forward-backward asymmetry of $B_c \to D_s^* \mu^+ \mu^-$ with $q^2$.

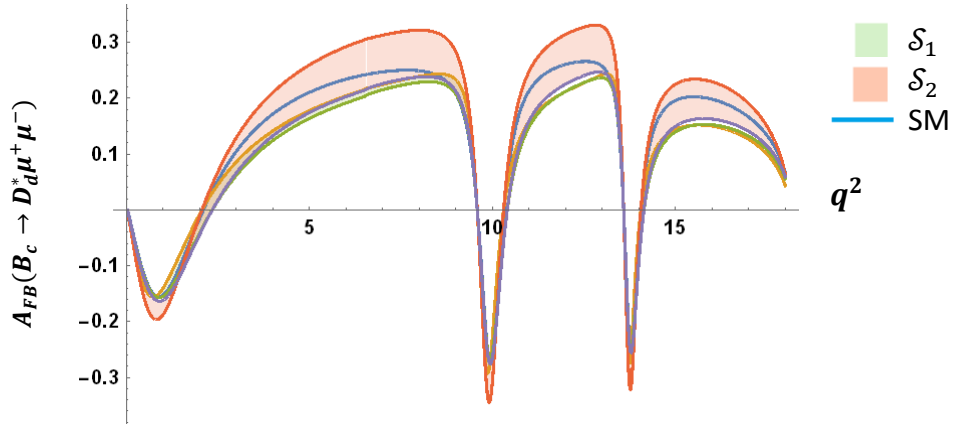

Fig. 14: The variation of forward-backward asymmetry of $B_c \to D_d^* \mu^+ \mu^-$ with $q^2$.

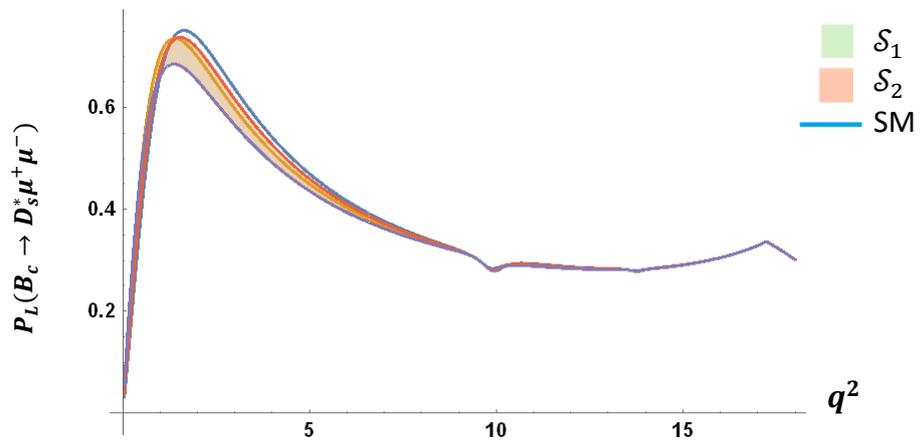

Fig. 15: The variation of polarization fraction of $B_c \to D_s^* \mu^+ \mu^-$ with $q^2$.

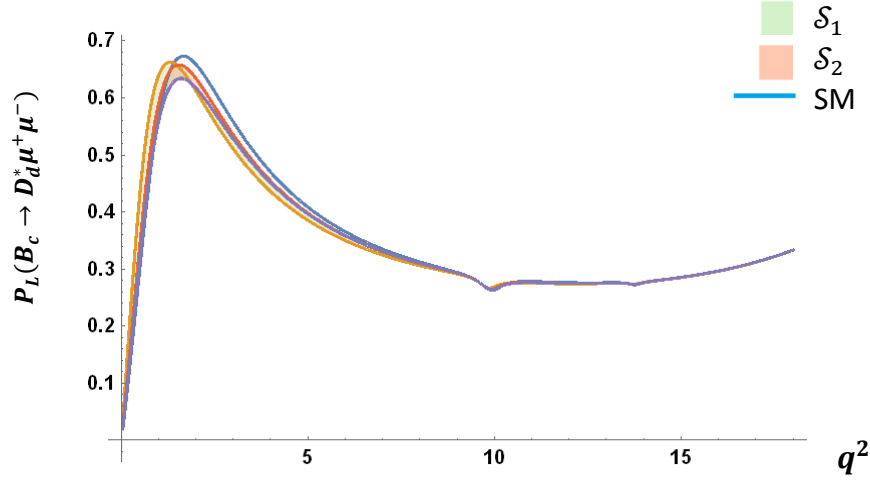

Fig. 16: The variation of polarization fraction of $B_c \to D_d^* \mu^+ \mu^-$ with $q^2$.

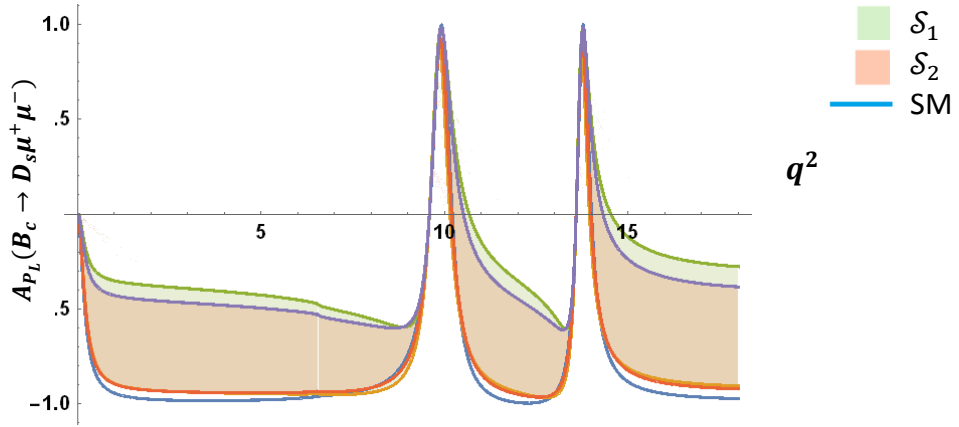

Fig. 17: The variation of lepton polarization asymmetry of $B_c \to D_s \mu^+ \mu^-$ with $q^2$.

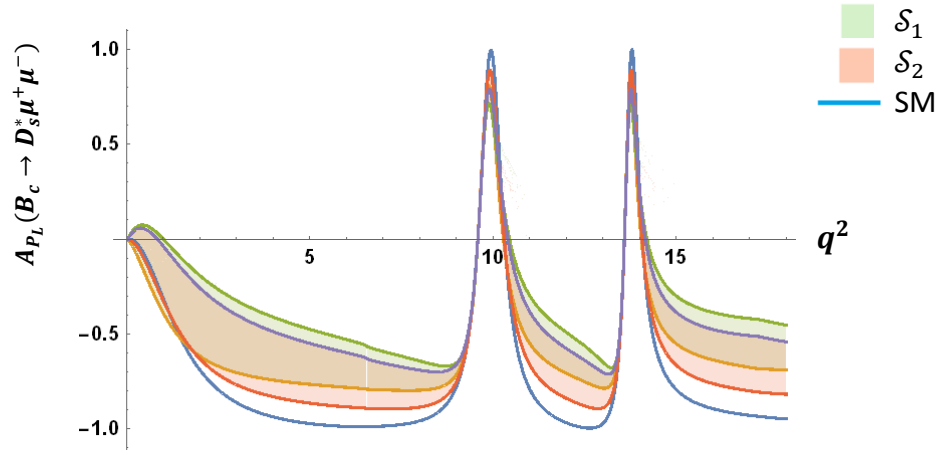

Fig. 18: The variation of lepton polarization asymmetry of $B_c \to D_s^* \mu^+ \mu^-$ with $q^2$.

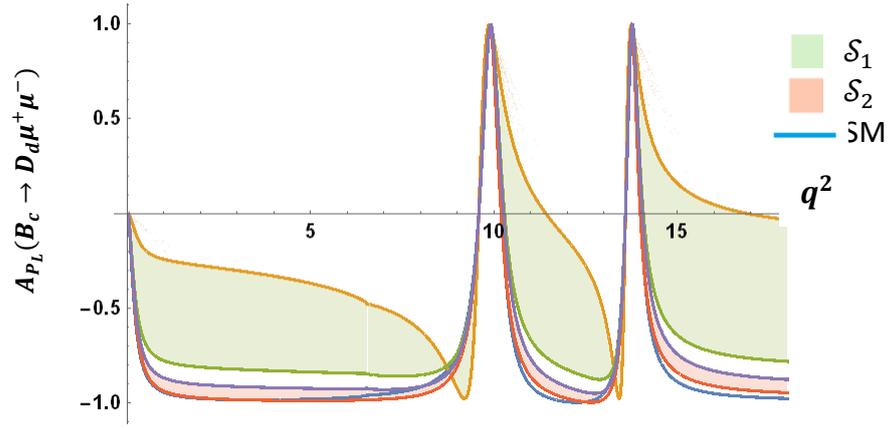

Fig. 19: The variation of lepton polarization asymmetry of $B_c \to D_d \mu^+ \mu^-$ with $q^2$.

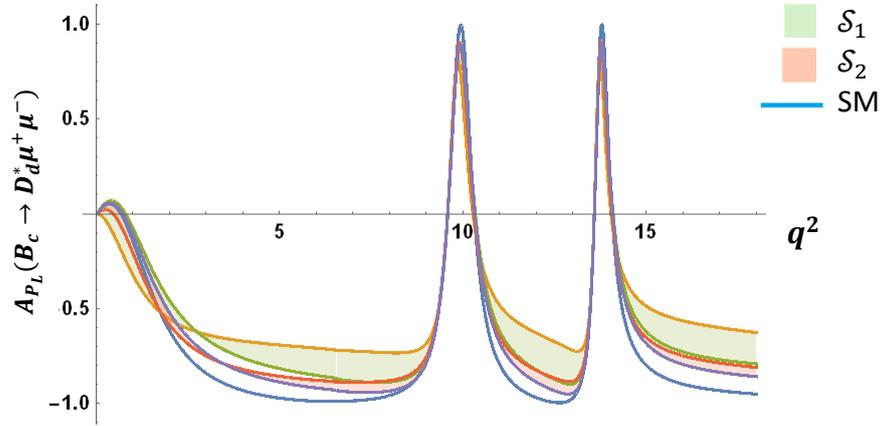

Fig. 20: The variation of lepton polarization asymmetry of $B_c \to D_d^* \mu^+ \mu^-$ with $q^2$.

Further we have plotted some correlation graphs between different decay observables and have shown them below. We have evaluated these graphs for low $q^2$ region i.e. $1 < q^2 < 6$ GeV$^2$.

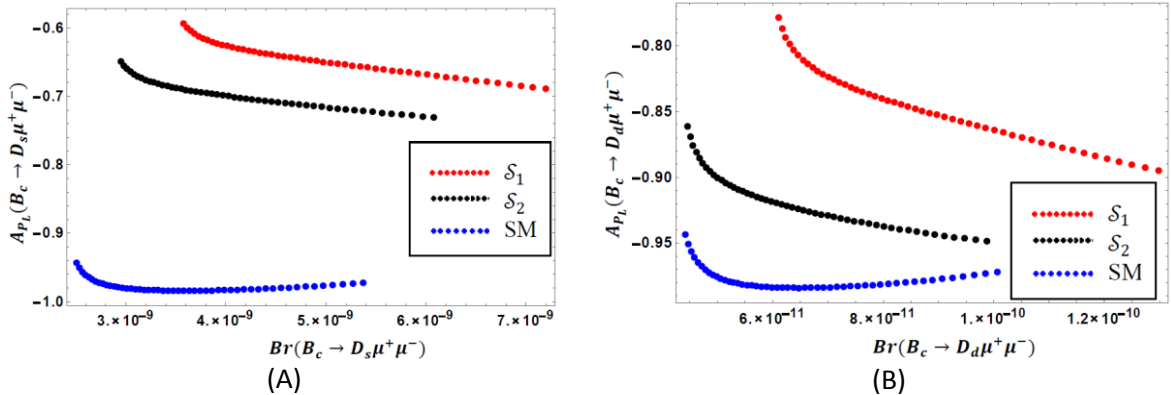

(A)            (B)

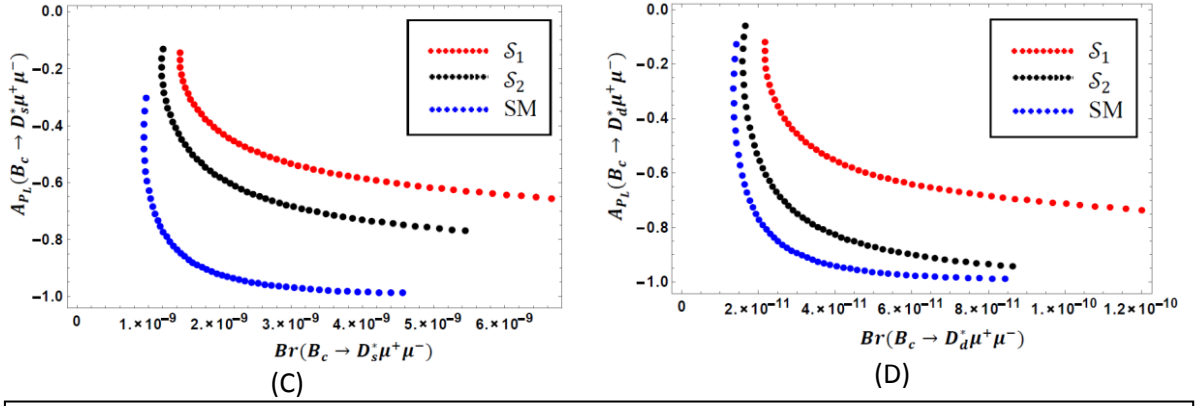

Fig. 21: Correlation between branching ratio ($Br$) and lepton polarization asymmetry ($A_{P_L}$) for : (A) $B_c \to D_s \mu^+ \mu^-$, (B) $B_c \to D_d \mu^+ \mu^-$, (C) $B_c \to D_s^* \mu^+ \mu^-$, (D) $B_c \to D_d^* \mu^+ \mu^-$ channels.

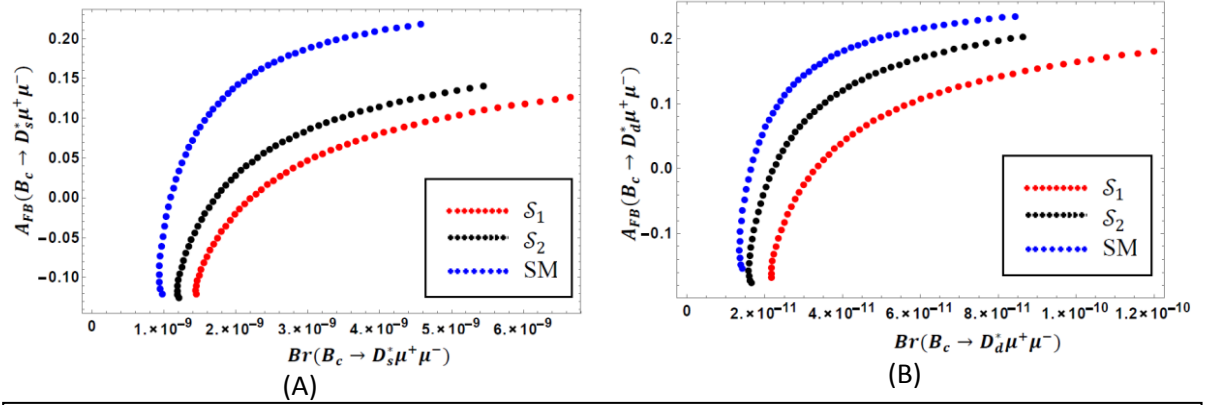

Fig. 22: Correlation between branching ratio ($Br$) and forward-backward asymmetry ($A_{FB}$) for: (A) $B_c \to D_s^* \mu^+ \mu^-$, (B) $B_c \to D_d^* \mu^+ \mu^-$ channels.

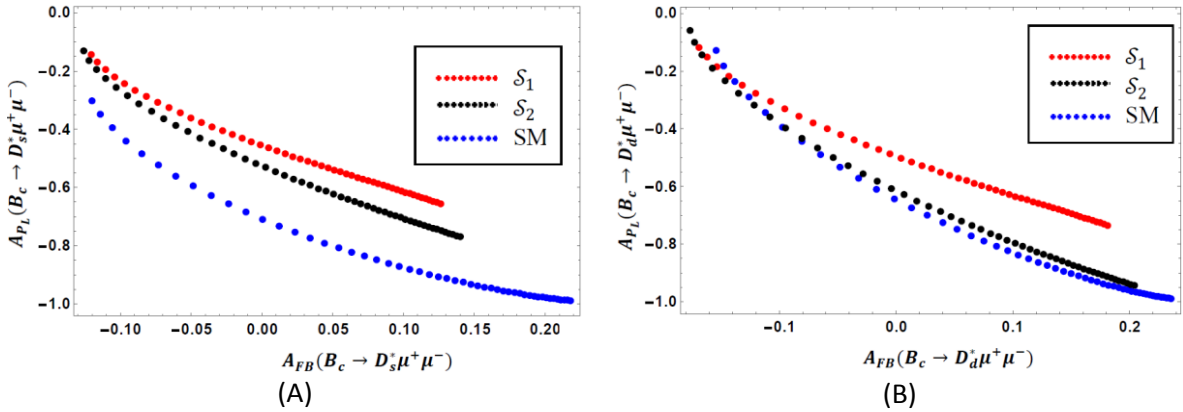

Fig. 23: Correlation between forward-backward asymmetry ($A_{FB}$) and lepton polarization asymmetry ($A_{P_L}$) for: (A) $B_c \to D_s^* \mu^+ \mu^-$, (B) $B_c \to D_d^* \mu^+ \mu^-$ channels.

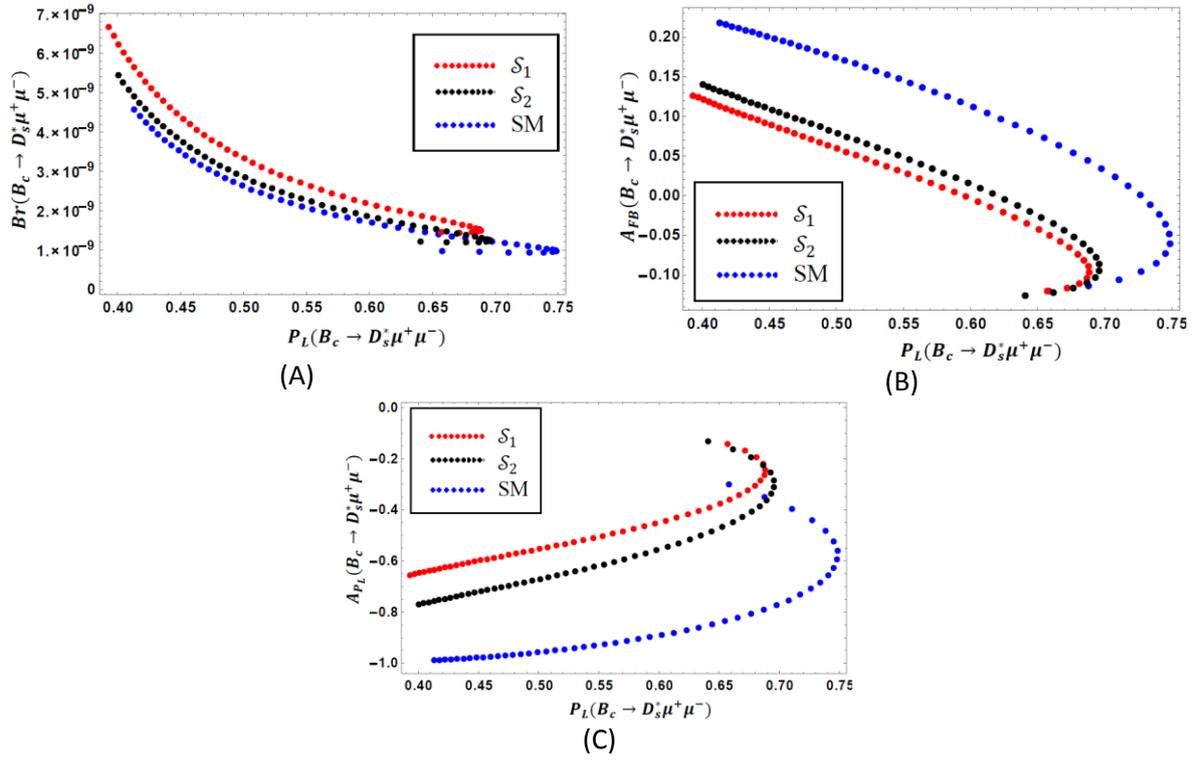

Fig. 24: Correlation plots of longitudinal polarization fraction ($P_L$) with: (A) branching ratio ($Br$), (B) forward-backward asymmetry ($A_{FB}$) and (C) lepton polarization asymmetry ($A_{P_L}$) for $B_c \to D_s^* \mu^+ \mu^-$ channel.

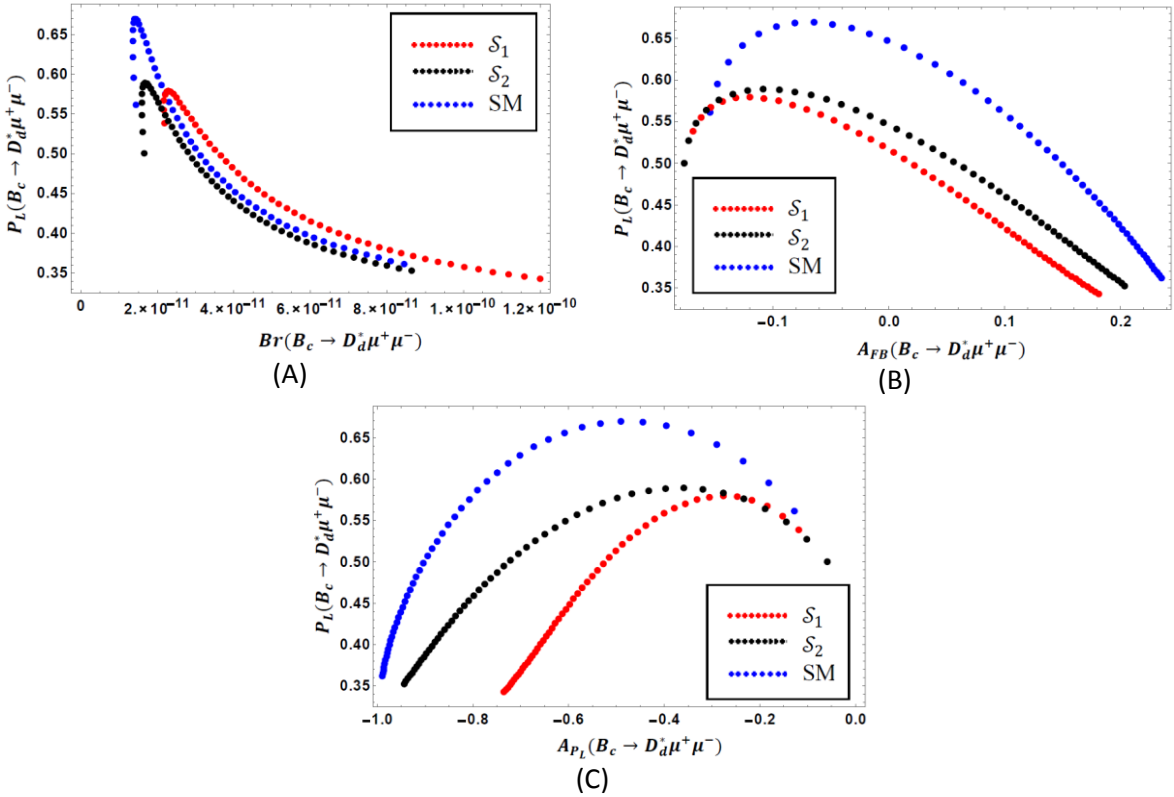

Fig. 25: Correlation plots of longitudinal polarization fraction ($P_L$) with: (A) branching ratio ($Br$), (B) forward-backward asymmetry ($A_{FB}$) and (C) lepton polarization asymmetry ($A_{P_L}$) for $B_c \to D_d^* \mu^+ \mu^-$ channel.

From the figures above we could gather some information about the dependence of different observables to each other. Fig. 21 (A, B, C, D) show that for both the scenarios of $Z'$ model, branching ratio and lepton polarization asymmetry enhance than their SM predictions. In $Z'$ model we found that the NP contribution to $B_c \to D_s^* \mu^+ \mu^-$ channel drives the polarization asymmetry, $A_{P_L}$, maximum up to $\sim -0.4$ which is much larger than the SM. Whereas the increment of $A_{P_L}$ is not that high but still above the SM values for other channels. These large deviations could hold remarkable NP signatures. Fig. 22 (A, B) would be important as it illustrates the dependence of forward-backward asymmetry to the branching ratio of $B_c \to D_s^* \mu^+ \mu^-$ and $B_c \to D_d^* \mu^+ \mu^-$ channels. Higher is the branching ratio, higher is the FB asymmetry and the correlation curves of both the scenarios of the NP model lie below the SM prediction in the low momentum transfer region. While plotting the correlation between FB asymmetry and lepton polarization asymmetry of $B_c \to D_s^* \mu^+ \mu^-$ and $B_c \to D_d^* \mu^+ \mu^-$ channels, we observed that the graphs for NP go above the SM and for higher FB asymmetry, the lepton polarization asymmetry decreases. The correlation graphs for longitudinal polarization fraction with other observables are shown in Figs. 24 (for $B_c \to D_s^* \mu^+ \mu^-$) and 25 (for $B_c \to D_s^* \mu^+ \mu^-$).

## VI. Conclusion

The decay modes $B_c \to D_{s,d}^{(*)} \mu^+ \mu^-$ are still in progress for experimental verification. In our analysis we have adopted the QCD motivated relativistic quark approach for the form factors. These form factors satisfy all the heavy quark and large energy symmetry relations which explicitly allow such transitions in the whole accessible kinematical range. Our study shows a noticeable effect of non-universal $Z'$ boson to these rare decay modes of $B_c$ meson which can be investigated in the LHCb experiment as $B_c$ mesons are expected to be produced numerously in near future.

The contribution of the cascade decays of $J/\psi(\psi)$ meson lying in the long-distance terms of matrix elements could be cut down in any experimental as well as theoretical approach as they provide huge uncertainties in decay distributions. From our study, we have got the cuts around $9 \text{GeV}^2 < q^2 < 15 \text{ GeV}^2$. The dependence of branching fraction to the $Z'$ model parameters shown here are basically for negative $D_{LL}$ and $S_{LL}$ as their large positive values are forbidden by the constraints from $A_{FB}(\bar{B}_d = K^* \mu^+ \mu^-)_{0<q^2<2 \text{ GeV}^2}$. All the branching fractions are increased in high $q^2$ region and the best increment shows up for scenario $S_1$. Other parameters such as FB asymmetry and lepton polarization asymmetry also increased in the NP model. The longitudinal polarization fraction of the vector mesons $D_{s,d}^*$ do not show much deviation from its SM value which states that $Z'$ boson would not change the polarization direction of those mesons. Also we have given some pictorial descriptions of correlation between different decay observables which might be worthy for further investigations of these decay channels. So as a concluding remark we can state that presence of non-universal $Z'$ boson provides a good rise in the decay rate as well as other decay observables of $B_c \to D_{s,d}^{(*)} \mu^+ \mu^-$ decay channels. This might be helpful for the search of these decay modes with highly sensitive experimental set up in future.

## Acknowledgement

We thank the reviewer for useful comments and suggestion which improve the quality of the paper. Maji is thankful to DST, Govt. of India for providing INSPIRE Fellowship (IF160115). Nayek and Sahoo are grateful to SERB, DST, Govt. of India for financial support through project (EMR/2015/000817). Mahata and Biswas thank NIT Durgapur for providing fellowship.

# Appendix A

$$y_{pert}(q^2) = h\left(\frac{m_c}{m_b}, \frac{q^2}{m_b^2}\right)(3C_1 + C_2 + 3C_3 + C_4 + 3C_5 + C_6) - \frac{1}{2}h\left(1, \frac{q^2}{m_b^2}\right)(4C_3 + 4C_4 + 3C_5 +$$
$$C_6) - \frac{1}{2}h\left(0, \frac{q^2}{m_b^2}\right)(C_3 + 3C_4) + \frac{2}{9}(3C_3 + C_4 + 3C_5 + C_6) \quad (A1)$$

where

$$h(z, s) = -\frac{8}{9}\ln(z) + \frac{8}{27} + \frac{4}{9}x - \frac{2}{9}(2 + x)\sqrt{|1 - x|}\begin{cases}\ln\left|\frac{\sqrt{1-x}+1}{\sqrt{1-x}-1}\right| - i\pi, & x \equiv \frac{4z^2}{s'} < 1 \\ 2\arctan\frac{1}{\sqrt{x-1}}, & x \equiv \frac{4z^2}{s'} > 1\end{cases}$$

$$h(0, s') = \frac{8}{27} - \frac{8}{9}\ln\frac{m_b}{\mu} - \frac{4}{9}\ln(s') + \frac{4}{9}i\pi. \quad (A2)$$

Here, $z = \frac{m_c}{m_b}, s = \frac{q^2}{m_b^2}$.

$$y_{BW}(q^2) = \frac{3\pi}{\alpha^2}\sum_{V_i=J/\psi,\psi}\frac{\Gamma(V_i \to l^+l^-)m_{V_i}}{m_{V_i}^2 - q^2 - im_{V_i}\Gamma_{V_i}}, \quad (A3)$$

The hadronic matrix elements for $B_c \to D_{s(d)}\mu^+\mu^-$ decays could be written in terms of three invariant meson to meson transition form factors. These are

$$\langle D_{s(d)}|\bar{s}\gamma^\mu b|B_c\rangle = f_+(q^2)\left[p_{B_c}^\mu + p_{D_{s(d)}}^\mu - \frac{M_{B_c}^2 - M_{D_{s(d)}}^2}{q^2}q^\mu\right] + f_0(q^2)\frac{M_{B_c}^2 - M_{D_{s(d)}}^2}{q^2}q^\mu,$$
$$\langle D_{s(d)}|\bar{s}\sigma^{\mu\nu}q_\nu b|B_c\rangle = \frac{if_T(q^2)}{M_{B_c}+M_{D_{s(d)}}}\left[q^2\left(p_{B_c}^\mu + p_{D_{s(d)}}^\mu\right) - \left(M_{B_c}^2 - M_{D_{s(d)}}^2\right)q^\mu\right]. \quad (A4)$$

Similarly, for $B_c \to D_{s(d)}^*\mu^+\mu^-$ channels the hadronic matrix elements could be parameterized in terms of seven invariant form factors. These are

$$\langle D_{s(d)}^*|\bar{s}\gamma^\mu b|B_c\rangle = \frac{2iV(q^2)}{M_{B_c} + M_{D_{s(d)}^*}}\epsilon^{\mu\nu\rho\sigma}\epsilon_\nu^* p_{B_c\rho} p_{D_{s(d)}^*\sigma},$$

$$\langle D_{s(d)}^*|\bar{s}\gamma^\mu\gamma_5 b|B_c\rangle = 2M_{D_{s(d)}^*}A_0(q^2)\frac{\epsilon^*\cdot q}{q^2}q^\mu + \left(M_{B_c} + M_{D_{s(d)}^*}\right)A_1(q^2)\left(\epsilon^{*\mu} - \frac{\epsilon^*\cdot q}{q^2}q^\mu\right)$$
$$-A_2(q^2)\frac{\epsilon^*\cdot q}{\left(M_{B_c} + M_{D_{s(d)}^*}\right)}\left[p_{B_c}^\mu + p_{D_s^*}^\mu - \frac{M_{B_c}^2 - M_{D_{s(d)}^*}^2}{q^2}q^\mu\right],$$

$$\langle D_{s(d)}^*|\bar{s}\sigma^{\mu\nu}q_\nu b|B_c\rangle = 2T_1(q^2)\epsilon^{\mu\nu\rho\sigma}\epsilon_\nu^* p_{B_c\rho} p_{D_{s(d)}^*\sigma},$$

$$\langle D_{s(d)}^*|\bar{s}\sigma^{\mu\nu}\gamma_5 q_\nu b|B_c\rangle = T_2(q^2)[\left(M_{B_c}^2 - M_{D_{s(d)}^*}^2\right)\epsilon^{*\mu} - (\epsilon^*\cdot q)(p_{B_c}^\mu + p_{D_{s(d)}^*}^\mu)]$$
$$+T_3(q^2)(\epsilon^*\cdot q)[q^\mu - \frac{q^2}{M_{B_c}^2 - M_{D_{s(d)}^*}^2}(p_{B_c}^\mu + p_{D_{s(d)}^*}^\mu)], \quad (A5)$$

where, $q^\mu = (p_B - p_{D_{s(d)}}, p_{D_{s(d)}^*})^\mu$ is the four momentum transfer and $\epsilon_\mu$ is polarization vector of $D_{s(d)}^*$ meson.

The helicity amplitudes for $B_c \to D_{s(d)} l^+ l^-$ decay mode are written as

$$H_\pm^{(i)} = 0,$$

$$H_0^{(1)} = \sqrt{\frac{\lambda}{q^2}} \left[ C_9^{eff} f_+(q^2) + C_7^{eff} \frac{2m_b}{M_{B_c} + M_{D_{s(d)}}} f_T(q^2) \right],$$

$$H_0^{(2)} = \sqrt{\frac{\lambda}{q^2}} C_{10} f_+(q^2),$$

$$H_t^{(1)} = \frac{M_{B_c}^2 - M_{D_{s(d)}}^2}{q^2} C_9^{eff} f_0(q^2),$$

$$H_t^{(2)} = \frac{M_{B_c}^2 - M_{D_{s(d)}}^2}{q^2} C_{10} f_0(q^2). \tag{A6}$$

Similarly for $B_c \to D_{s(d)}^* l^+ l^-$ modes, the hadronic helicity amplitudes are

$$H_\pm^{(1)} = -\left(M_{B_c}^2 - M_{D_{s(d)}^*}^2\right)\left[ C_9^{eff} \frac{A_1(q^2)}{\left(M_{B_c} - M_{D_{s(d)}^*}\right)} + \frac{2m_b}{q^2} C_7^{eff} T_2(q^2) \right]$$
$$\pm \sqrt{\lambda} \left[ C_9^{eff} \frac{V(q^2)}{\left(M_{B_c} + M_{D_{s(d)}^*}\right)} + \frac{2m_b}{q^2} C_7^{eff} T_1(q^2) \right],$$

$$H_\pm^{(2)} = C_{10} \left[ -\left(M_{B_c} + M_{D_{s(d)}^*}\right) A_1(q^2) \right] \pm \frac{\sqrt{\lambda}}{\left(M_{B_c} + M_{D_{s(d)}^*}\right)} C_{10} V(q^2),$$

$$H_0^{(1)} = -\frac{1}{2 M_{D_{s(d)}^*} \sqrt{q^2}} \Bigg\{ C_9^{eff} \left[ \left(M_{B_c}^2 - M_{D_{s(d)}^*}^2 - q^2\right)\left(M_{B_c} + M_{D_{s(d)}^*}\right) A_1(q^2) - \frac{\lambda}{M_{B_c} + M_{D_{s(d)}^*}} A_2(q^2) \right] +$$
$$2 m_b C_7^{eff} \left[ \left(M_{B_c}^2 + 3 M_{D_{s(d)}^*}^2 - q^2\right) T_2(q^2) - \frac{\lambda}{M_{B_c}^2 - M_{D_{s(d)}^*}^2} T_3(q^2) \right] \Bigg\},$$

$$H_0^{(2)} = -\frac{1}{2 M_{D_{s(d)}^*} \sqrt{q^2}} C_{10} \left[ \left(M_{B_c}^2 - M_{D_{s(d)}^*}^2 - q^2\right)\left(M_{B_c} + M_{D_{s(d)}^*}\right) A_1(q^2) - \frac{\lambda}{M_{B_c} + M_{D_{s(d)}^*}} A_2(q^2) \right],$$

$$H_t^{(1)} = \sqrt{\frac{\lambda}{q^2}} C_9^{eff} A_0(q^2),$$

$$H_t^{(2)} = \sqrt{\frac{\lambda}{q^2}} C_{10} A_0(q^2), \tag{A7}$$

where,

$$\lambda = M_{B_c}^4 + M_{D_{s(d)}^{(*)}}^4 + q^4 - 2\left(M_{B_c}^2 M_{D_{s(d)}^{(*)}}^2 + M_{D_{s(d)}^{(*)}}^2 q^2 + M_{B_c}^2 q^2\right).$$